\documentstyle[prl,aps,epsfig]{revtex}         

\catcode`\ä = \active \catcode`\ö = \active \catcode`\ü = \active
\catcode`\Ä = \active \catcode`\Ö = \active \catcode`\Ü = \active
\catcode`\ß = \active \catcode`\é = \active \catcode`\è = \active
\catcode`\ë = \active \catcode`\ô = \active \catcode`\ê = \active
\catcode`\ø = \active \catcode`\ò = \active
\defä{\"a}
\defö{\"o}
\defü{\"u}
\defÄ{\"A}
\defÖ{\"O}
\defÜ{\"U}
\defß{\ss}
\defé{\'{e}}
\defè{\`{e}}
\defë{\"{e}}
\defô{\^{o}}
\defê{\^{e}}
\defø{\o}
\defò{\`{o}}
\begin{document}
\draft               
\title{Dissipationless flow and superfluidity in gaseous Bose-Einstein
condensates} \author{C. Raman, R. Onofrio, J. M. Vogels, J. R.
Abo-Shaeer, and W. Ketterle} \address{Department of Physics and
Research Laboratory of Electronics, \\
Massachusetts Institute of Technology, Cambridge, MA 02139}
\date{\today{}}
\maketitle

\begin{abstract}
We study dissipation in a dilute Bose gas induced by the motion of a macroscopic object. A
blue-detuned laser beam focused on the center of a trapped gas of sodium atoms was scanned both
above and below the BEC transition temperature.  The measurements allow for a comparison between
the heating rates for the superfluid and normal gas.
\end{abstract}


\section{INTRODUCTION}

Superfluid flow is a manifestation of quantum mechanics at the macroscopic level.  Phenomena like
dissipationless flow and persistent currents can be traced back to the existence of a macroscopic
wavefunction.  The gradient of its phase gives the superfluid velocity \cite{nozi90}.  However,
superfluid flow is only stable below a critical velocity.  Above this velocity, excitations of the
fluid can be generated, which can be either phonons, rotons (in liquid $^4$He) or vortices.

The realization of Bose-Einstein condensation in dilute gases
\cite{kett99var,corn99var} has created a new test ground for the
theory of quantum fluids.  Early experiments confirmed the
microscopic foundation of the phenomenon of superfluidity: the
phonon nature of low-lying collective excitations was observed
\cite{jin96coll,mewe96coll,andr97prop} and the coherence or
macroscopic phase of the condensate \cite{andr97int}.  This
picture became more complete during the last year with the
observation of vortices \cite{matt99vort,madi00}, transverse
excitations \cite{mara00scis}, evidence for a critical velocity
\cite{rama99} and suppression of collisions \cite{chik00}.
Although the metastability and therefore limited lifetime of the
quantum gas and its mesoscopic size may prevent spectacular
observations of persistent flows, there is already a clear
picture emerging of the many facets of superfluidity in a gas.

In this article we present measurements which indicate a regime of frictionless flow for a gaseous
Bose-Einstein condensate.  We emulate the motion of a macroscopic object by scanning a focused,
blue-detuned laser beam through the gas at different velocities. By taking measurements on nearly
pure condensates as well as purely thermal ensembles, we can compare the relative role of
dissipation in both phases.  This extends our previous work \cite{rama99,onof00sup} where we
studied dissipation in a Bose condensed system.

\section{THE EXPERIMENTAL SET-UP}

\bigskip The experiments were performed in a new apparatus for the
production of Bose-Einstein condensates of sodium atoms.  The apparatus has been functioning since
January 1999 and incorporates some improvements with respect to the original machine of the MIT
group.  In particular, a Zeeman slower with magnetic field reversal ("spin-flip" Zeeman slower)
delivered a flux of typically $10^{11}$ atoms s$^{-1}$.  This resulted in trapping $2-3 \times
10^{10}$ atoms in a dark-spontaneous force optical trap at a temperature of $\simeq$ 1 mK. After
3-4 ms of polarization gradient cooling, atoms at a temperature of 50-100 $\mu$K were loaded into
an Ioffe-Pritchard magnetic trap.  The trap consisted of four Ioffe bars and two elongated pinch
coils, symmetrically located around a quartz cell, combining tight confinement and excellent
optical access on four sides of the glass cell.  Under the tightest confinement, the cylindrically
symmetric confining potential has radial and axial trapping frequencies $\nu_r = 547$ Hz and
$\nu_z = 26$ Hz.  The initial stage of evaporative cooling (18 s) proceeded using this trapping
configuration; however, the final stage evaporation appeared to be limited by inelastic losses at
high densities. Therefore, the magnetic confinement was reduced in the last 5-7 s, finally
producing condensates of 2-5$\times 10^7$ atoms in traps of radial frequencies $\nu_r = 40-80$ Hz
and axial frequency $\nu_z = 20$ Hz.  Since decompression occurred during the evaporation and not
after the condensate is produced, as in previous work \cite{rama99}, non-adiabatic effects which
lead to heating were eliminated. Thus nearly pure condensates ($> 90$\% condensate fraction) were
produced regardless of the final trapping geometry.

\begin{figure}[htbp]
\epsfxsize=85mm \centerline{\epsfbox{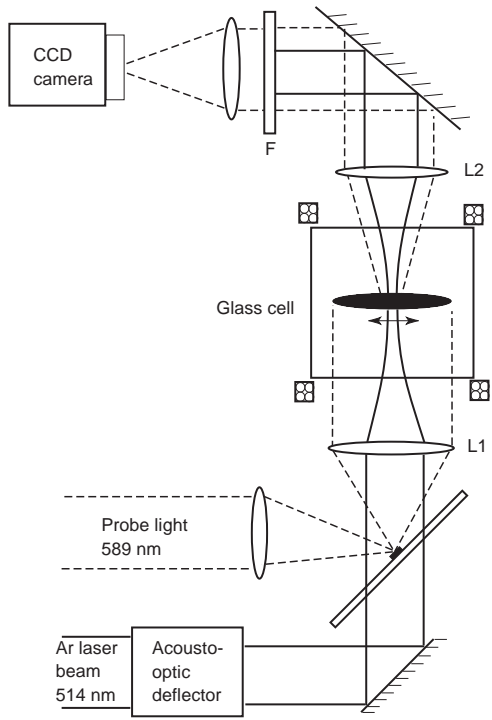}} \vspace{0.3cm}
\caption{Schematic view of the experimental setup. The condensate
was confined in a glass cell using an Ioffe-Pritchard magnetic
trap. The blue-detuned laser beam was sent through a two-axis AOM
scanner and then focused on the condensate using the lens L1. The
probe light was overlapped using a small mirror and, after passing
through the condensate, was focused with the lens L2 and sent to
the CCD camera. The filter F was used to avoid exposure of the
CCD camera to the high intensity blue-detuned laser beam.}
\label{setup}
\end{figure}

A 514 nm laser beam used to stir the atoms was split off from the
argon-ion laser that pumped the dye laser used for the cooling and
trapping beams.  Such focused blue-detuned beams have been used
previously to eliminate Majorana losses in the center of a
quadrupole magnetic trap \cite{davi95bec} and to excite sound
waves \cite{andr97prop}. A single mode optical fiber
significantly improved the stability with respect to previous
experiments \cite{rama99}.  The position of the laser focus did
not change by more than a few $\mu$m per day and required only
minimal adjustment to keep it centered on the atomic cloud.  The
laser was focused to a Gaussian $ 1/e^{2}$ beam diameter of 10 to
13 $\mu$m.  The repulsive optical dipole force expelled the atoms
from the regions of highest laser intensity, and we typically
used a ratio of barrier height to chemical potential (or
temperature, in the case of purely thermal clouds) of 5-10. The
condensate had a mean-field energy of 60-140 nK, while the
thermal cloud temperature was varied between 1 and 6 $\mu$K.
Thus, about 200-400 $\mu$W of laser power were required for
stirring a condensate, and 6-10 mW for stirring a thermal cloud.

The laser was focused on the center of the cloud by observing the
"pierced" condensate using in-situ phase-contrast imaging
\cite{onof00sup}.  The focus was scanned back and forth along the
axial direction using an acousto-optic deflector.  The velocity
of the scan was varied by adjusting the frequency $f$ and keeping
the amplitude $\alpha$ fixed, yielding a velocity of $v = 2
\alpha f$. The scan amplitude was chosen to be between a few beam
diameters and an amplitude of one-half the diameter of the cloud,
about 80 $\mu$m.  For thermal clouds, we typically scanned the
full diameter of the cloud.  A schematic of the experimental
set-up, including the imaging system, is depicted in Fig.
\ref{setup}.

\section{ENERGY DISSIPATION IN BOSE-EINSTEIN CONDENSATES}

Exposing a Bose condensate to the stirring laser beam transferred
energy to the atoms, resulting in an increase in the thermal
component.  We measured this component by turning off the scan,
allowing the gas to equilibrate for 100-200 ms and then shutting
off the magnetic trap.  After ballistic expansion for a fixed
time, typically 50 ms, the atoms were exposed to a $\sim$ 500
$\mu$s pulse of near-resonant light, and the shadow cast by the
atoms was imaged onto a CCD camera. The two-dimensional
transmission profile yielded the column density of the cloud and
the kinetic energy distribution of the atoms.  For mixed clouds
this distribution was bimodal, with the condensate localized in
the center and the more dilute thermal component appearing in the
wings. Our technique allowed us to measure energy changes as
little as 10 nK, in a regime where the thermal fraction ($<10\%$)
was barely visible in the images.

We employed this technique in earlier work to analyze the heating
induced by stirring the condensate with a blue-detuned laser beam
at various velocities\cite{rama99}. There the temperature
increase suggested two regimes of dissipation separated by a
velocity threshold. However, a precise determination of the
critical velocity was not possible since the sensitivity of the
measurements was limited by fluctuations in the background heating
processes such as inelastic collisions. As a result, the small
heating rate due to the stirrer near the critical velocity, of
the order of 20 nK/s, could not be detected. Moreover, the
initial thermal fraction could not be suppressed below 40\% due
to non-adiabatic effects during decompression of the magnetic
trap.

By optimizing the evaporation strategy, we have achieved nearly pure ($> 90$\%) condensates as
initial conditions for the stirring experiments.  While background heating processes could not be
fully eliminated, we could account for them by subtracting the energy of stirred and unstirred
clouds. Thus, even with a nearly invisible thermal component we can detect changes in the
non-condensed fraction of a few percent. Using this method, we have measured the small heating
present near the critical velocity that was previously indiscernible.

\begin{figure}[tbp]
\begin{center}
\epsfxsize=100mm \centerline{\epsfbox{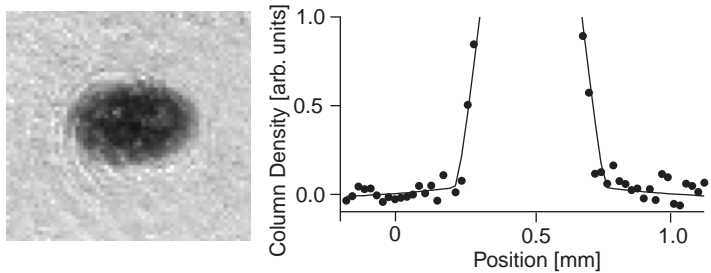}}
\end{center}
\caption{High sensitivity calorimetry.  Shown is an absorption image of an expanded cloud.  On the
right is a one-dimensional vertical slice taken through the center of the image which represents
the column density $\tilde{n} = -\ln{T}$, where $T$ is the transmission of the probe light. The
vertical axis is expanded to see the extremely dilute thermal wings. The solid line is a
constrained two-dimensional bimodal fit, which yields a thermal fraction of $(6 \pm 0.6)$ \%.}
\label{bimodal}
\end{figure}
Fig. \ref{bimodal} shows a normalized image of the probe light
transmitted by the atomic cloud. The data can be described by a
bimodal velocity distribution comprised of a Thomas-Fermi profile
for the condensate and a Gaussian distribution for the thermal
component \cite{kett99var},
\begin{equation}
\tilde{n}(\rho,z) = \tilde{n}_C
\left(1-\frac{\rho^2}{\rho_C^2}-\frac{z^2}{z_T^2}\right)^{3/2}+
\frac{\tilde{n}_T}{g_2(1)} g_2(e^{-(\rho^2/\rho_T^2+z^2/z_T^2)})
\label{eq:bimodal_fit}
\end{equation}
where $\rho$ and $z$ are the radial and axial directions,
respectively, $\rho_C$ and $\rho_T$ the radial size of condensate
(Thomas-Fermi radius) and thermal component, respectively, and
$z_C$ and $z_T$ the same for the axial sizes.  $\tilde{n}_C$ and
$\tilde{n}_T$ are the peak column density of the condensate and
thermal cloud in the time-of-flight absorption image.  The Bose
function is $g_n (x) = \sum_{k=1}^{\infty} x^k/k^n$.

The temperature of the thermal cloud is determined from the width
of the Gaussian distribution,
\begin{equation}
k_B T = \frac{1}{2} M z_T^2 \left(\frac{\omega_{z
}^2}{1+\omega_{z}^2t^2}\right)\simeq \frac{1}{2} M
\frac{z_T^2}{t^2}
\end{equation}
where the latter approximation holds for long time of flight
$\omega_z t >> 1$. The number of thermal atoms is $N_T = \int
\tilde{n}_T(\rho,z)dz = \pi [g_3(1) /g_2(1)] \tilde{n}_T \rho_T
z_T \simeq 2.3\; \tilde{n}_T \rho_T z_T$. Due to the high optical
density in the center of the images, the number of condensate
atoms was more reliably extracted from the mean-field energy
rather than the integrated column density.  In the Thomas-Fermi
limit this is $\mu = \frac{1}{2} (15 \bar{\omega}^3 \hbar^2
\sqrt{M} N_C a)^{2/5}$, where $\bar{\omega} = (\omega_r^2
\omega_z)^{1/3}$ is the mean angular trapping frequency, and $a =
2.75$ nm is the two-body scattering length for sodium. Equating
mean-field energy to the measured kinetic energy we get $\mu =
\frac{1}{2} M (\rho_C^2+\frac{1}{2} z_C^2)/t^2$, where $t$ is the
time-of-flight. The total number is $N=N_T+N_C$.

The transmitted fraction of the probe light was determined by the
ratio of images taken with and without the atoms present.  The
two-dimensional data were fitted to the function $f(\rho,z) =
Ce^{-\sigma \tilde{n}(\rho,z)}$, where $\sigma$ is the known
absorption cross-section.  In general, one can fit all of the
unknown parameters $N_T/N,N_C/N,N,\rho_T,z_T,\rho_C,z_C$, as well
as the rotation angle of the image in the $\rho$-$z$ plane and a
scale factor $C$ caused by fluctuations in the probe light
intensity between the two images.

For very low temperatures the spatial extent $\rho_T,z_T$ of the
thermal component as well as its column density $\tilde{n}_T$
diminish considerably. It became very difficult to reliably
extract thermal fractions below 10\% from an unconstrained
bimodal fit of the form given in Eqn. \ref{eq:bimodal_fit}.
Therefore, we constrained the thermal fit parameters
$N_T/N,\rho_T$ and $z_T$ to reflect the correlation between the
number of thermally excited atoms and the temperature, a
relationship which may be parameterized in different ways.  For
example, for an ideal Bose gas, the thermal fraction and energy
per particle are
\begin{eqnarray}
{\left.\frac{N_T}{N} \right |}_0 &=& \left(\frac{T}{T_c} \right)^3  \\
{\left.\frac{E}{N k_B T_c} \right |}_0 &=& \frac{3 g_4(1)}{g_3(1)}
\left (\frac{T}{T_c} \right )^4
\end{eqnarray}
where $T_{c}$ is the transition temperature. Including the effect
of the finite mean-field energy $\mu$ increases the thermal
fraction (and therefore the total energy), yielding a correction
to the above expressions
\begin{eqnarray}
\frac{N_T}{N} &=& {\left.\frac{N_T}{N} \right |}_0 +
\xi(T/T_{c},\eta) \label{eq:thermo1} \\ \frac{E}{Nk_B T_{c}} &=&
{\left.\frac{E}{N k_B T_{c}} \right |}_0 +\tau(T/T_{c},\eta)
\label{eq:thermo2}
\end{eqnarray}
where $\eta = \mu_{T=0}/k_BT_{c}$.  We parameterized the
functions $\xi$ and $\tau$ which had been numerically evaluated
for specific values of $\eta$ in the range 0.3-0.45
\cite{gior97jltp} using mean-field theory. By inverting equation
\ref{eq:thermo1} we obtain for the total energy of the gas

\begin{equation}
\frac{E}{Nk_B T_{c}} = \epsilon(N_T/N,\eta)
\end{equation}
where the function $\epsilon$ is determined by the functions
$\xi$ and $\tau$.  The parameters $n_T, \rho_T$, and $z_T$ can
now be expressed in terms of a single parameter, the thermal
fraction $N_T/N$.  Without this constraint, the fitting routine
tended to increase the spatial extent $\rho_T,z_T$ of the thermal
cloud when the column density $\tilde{n}_T$ is low, an unphysical
result.

By imposing this constraint we could extend the bimodal fitting technique to temperatures close to
the chemical potential $\mu$.  Fig. \ref{bimodal} shows a one-dimensional vertical slice taken
through the image data, converted into column density of the cloud. The thermal wings appear at
about $0.3$ mm from the center of the image, near the edge of the condensate. Although the thermal
cloud is extremely dilute, the constrained fitting routine has no difficulty detecting the $6\%$
normal fraction of the gas present in the image.  With this technique, the energy per particle of
the gas can be established with an uncertainty of about 10 nK, for temperatures as low as $k_B
T/\mu = 1.2$. To our knowledge, this combination of high sensitivity and low temperature
thermometry has not been previously reported.

\begin{figure}[tbp]
\begin{center}
\epsfxsize=100mm \centerline{\epsfbox{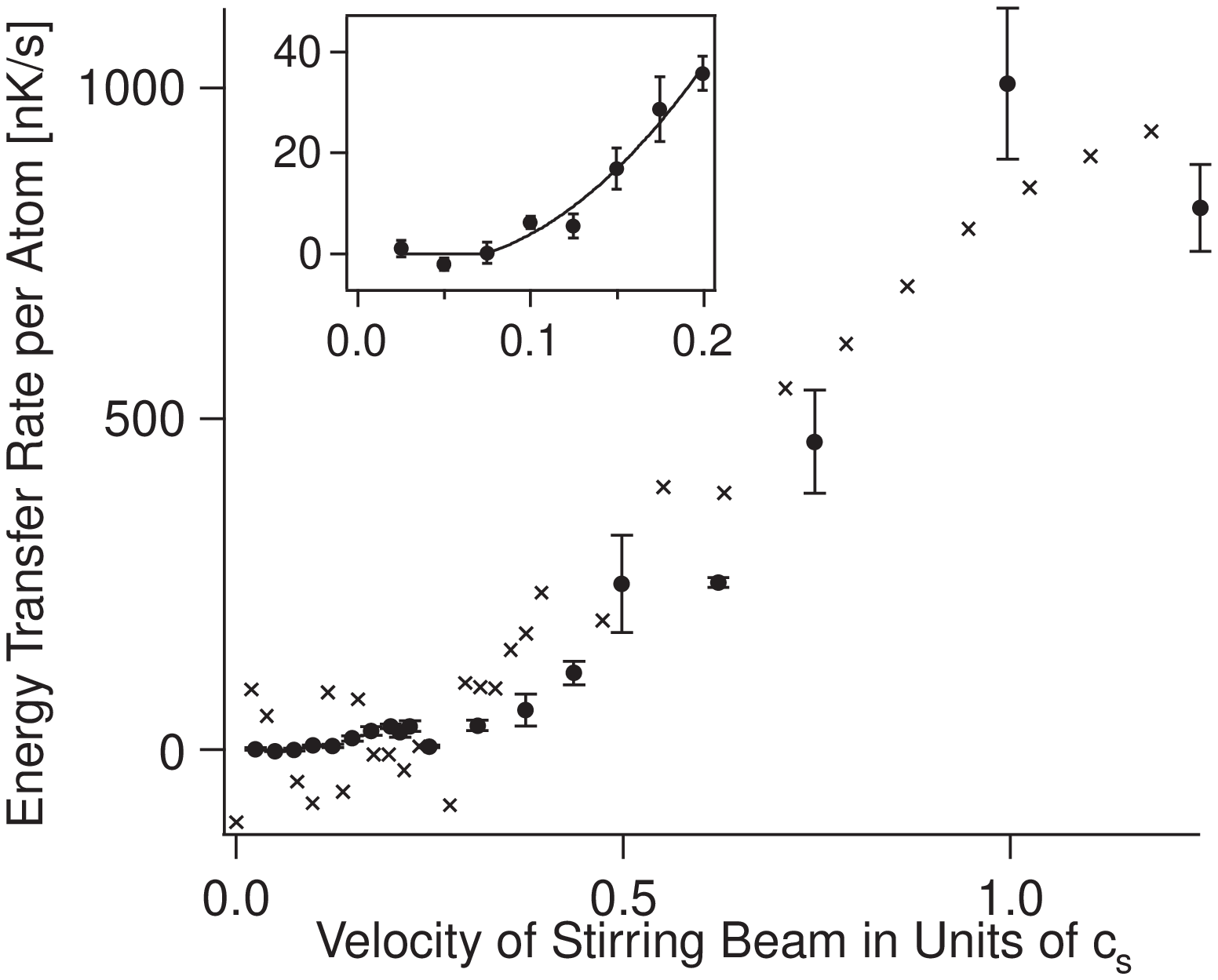}}
\vspace{0.3cm}
\end{center}
\caption{Calorimetry of a condensate. Shown is the energy
transfer rate versus velocity of the stirring laser beam measured
using subtraction of background heating processes (full circles)
and older data taken from ref. {\protect \cite{rama99}} without
background subtraction (crosses). The velocity is in units of the
peak sound velocity in the condensate. The error bars reflect
shot-to-shot variations in the temperature, which is the largest
source of error.  The inset is a magnification of the region near
zero velocity, showing sensitivity to $\sim 20$ nK/s heating
rates.} \label{condensate_calorimetry}
\end{figure}

Fig. \ref{condensate_calorimetry} shows the heating of the
condensate versus the velocity of the laser beam.  The velocity
was varied by changing the frequency of the laser beam while
keeping the scan amplitude fixed at roughly 1/3 of the axial
Thomas-Fermi diameter.  The heating rate was measured by
adjusting the stirring time between 30 ms and 8 s while keeping
the total energy transfer roughly the same for all velocities
used. Earlier data taken by varying the amplitude while keeping
the frequency fixed at 56 Hz, about 3 times the axial trapping
frequency \cite{rama99}, are shown for comparison.

The new data show a substantial improvement in signal-to-noise
with respect to the older set. This enhanced sensitivity allows
for a clear resolution of the small heating rate ($\simeq 20$
nK/s) visible at low velocities near $v/c_s \simeq 0.15$.  This
is shown in the inset, which is a magnified plot of the data in
the low velocity region.  These data suggest that the critical
velocity for excitation occurs at $v_c \simeq 0.1 c_s$, as
confirmed by direct measurements of the drag force from the
asymmetry of the condensate density profile induced by the
scanning laser beam \cite{onof00sup}. This technique was shown to
be consistent with the calorimetric measurements, but had a
higher sensitivity at low velocities.

Above the critical velocity the condensate experiences a drag
force $F = \kappa (v-v_c)$, where $\kappa$ is the coefficient of
drag \cite{fris92,jack99}. The rate of energy transfer is $F
\cdot v$, and so we can find the critical velocity from a
parabolic fit above $v_c$,
\begin{equation}
\frac{dE}{dt} = \kappa v(v-v_c)
\label{eq:fit}
\end{equation}
This yields a critical velocity of $v_c/c_s = 0.07 \pm 0.01$ for the new data. This new value is
smaller than estimated in our earlier paper \cite{rama99}. Both experiments are consistent, since
the scatter in the observed heating rates in \cite{rama99} prevented a clear observation of the
threshold, and the critical velocity was estimated by linear extrapolation from high heating rates
to be $v_c/c_s=0.25$. Other possible extrapolations tend to lower the value of $v_c$, for instance
a function $\propto v(v-v_c)$ as suggested by equation \ref{eq:fit}, yielded $v_c/c_s=0.20 \pm
0.07$. With our improved calorimetry we have done measurements using the fixed frequency/variable
amplitude method employed earlier and observed a small heating rate of $20$ nK/s at $v = 0.15c_s$
that was previously indiscernible. Furthermore, small differences in the laser beam profile may
have also contributed to the higher value of $v_c$ in \cite{rama99}.

Both data sets show good agreement in the overall heating rate.
For the new data above $v/c_s \simeq 0.1$ the heating rate
increases, until about $0.2$, where it appears to level off and
then to drop to a minimum at around $v/c_s\simeq 0.25$. For
higher velocities the heating rate increases once again.  The
suppression of the heating occurring at $v/c_s \simeq 0.25$ arises
from a frequency dependent feature, as discussed in
\cite{onof00sup}. At that velocity the laser scans near the axial
trapping frequency and excites synchronous dipole motion of the
condensate.  This results in a lower relative velocity between
condensate and stirrer, thereby reducing the heating.

\section{ENERGY DISSIPATION IN A NORMAL GAS}

At finite temperature, one would expect heating even below the
critical velocity due to physical processes occurring within the
normal component. For a dilute Bose gas at temperatures below 0.45
$T_{c}$, the thermal fraction is less than 10\%, and plays a very
minor role in the heating discussed so far.  In addition, the
repulsive mean field of the condensate lowers the density of the
normal component in the center of the cloud even further.  This
allowed us to study the breakdown of superfluidity due to
processes initiated primarily within the condensate.

There is a a major difference in the description of the condensate and the normal component.
Interactions within the condensate give rise to quantum fluid properties. The macroscopic
wavefunction $\Psi = \sqrt{n} e^{i\phi}$ yields hydrodynamic equations of motion for this fluid,
expressed in terms of the density $n$ and superfluid velocity ${\mathbf{v}} = (\hbar/m) \nabla
\phi$. In contrast, a dilute normal gas interacts very weakly with itself and with the condensate.
For the most part, the mean free path for collisions $l_{\rm mfp}> d$, where $d$ is some
characteristic length scale, for example, the laser beam diameter or the radius of the cloud.  In
this collisionless regime, no correlations between particles exist, and a fluid description in
terms of a local density $n$ and a local velocity $\mathbf{v}$ does not hold. However, heating
still occurs as long as the thermal atoms collide with the moving laser beam. We present
measurements of this heating mechanism in a gas above the transition temperature, which we can
understand using a kinetic model. These measurements allow us to compare the relative
effectiveness of heating the normal and superfluid components.

\begin{figure}[tbp]
\begin{center}
\epsfxsize=50mm \centerline{\epsfbox{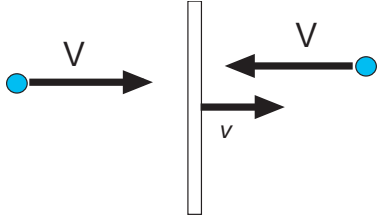}}
\vspace{0.3cm}
\end{center}
\caption{One-dimensional model for the heating of a thermal
cloud. Atoms from both sides collide at velocity $V$ with a wall
moving at velocity $v$.} \label{moving_wall}
\end{figure}

Let us first consider a one-dimensional model of a hard wall
moving at a velocity $v$ through a gas of atoms with thermal
velocity $V \sim \sqrt{2 k_B T/M}$ (see Fig. \ref{moving_wall}).
We assume $v\ll V$ and elastic collisions between the atoms and
the wall.

An atom moving in the direction of the wall slows down during the
collision, while an atom moving against the wall increases its
velocity. This leads to a net heating rate due to the finite
velocity $v$. The energy transferred to atoms on the left and
right is
\begin{eqnarray}
\Delta E_{\rm left} &=&{\frac{1}{2}}M(-V+2v)^{2}-{\frac{1}{2}}MV^{2}=2Mv(v-V) \\
\Delta E_{\rm right} &=&{\frac{1}{2}}M(V+2v)^{2}-{\frac{1}{2}}MV^{2}=2Mv(v+V)
\end{eqnarray}
while the corresponding collision rates are
\begin{eqnarray}
\Gamma _{\rm left}&=&nA(V-v) \\
\Gamma _{\rm right}&=&nA(V+v)
\end{eqnarray}
where $n$ is the atomic density and $A$ the surface area of the
moving wall. The rate of energy increase is therefore
\begin{equation}
{\frac{dE}{dt}}=\Gamma _{\rm left}\Delta E_{\rm left}+\Gamma _{\rm right}\Delta E_{\rm
right}=8MnAVv^{2}
\end{equation}
which can be simply interpreted as an energy transfer per
collision, $\propto Mv^{2}$, multiplied by the collision rate
$nAV$. The model is easily extended to three dimensions and
objects of arbitrary shapes.  We approximate the stirring laser
as an infinite cylinder of diameter $d$ moving transversely to
its axis of symmetry, along the axial direction of the
cylindrically symmetric trap.  The frequencies of oscillation
along the axial and radial direction are $\omega_{z}$ and
$\omega_{r}$, respectively, and the cloud $1/e^2$ radii are
$R_{r,z} = \sqrt{2k_B T /M \omega_{r,z}^2}$.  Summing over the
Boltzmann velocity distribution, and accounting for the
three-dimensional nature of the collisions, the energy transfer
rate per particle is:

\begin{equation}
{\left .\frac{dE}{dt} \right |}_{N}= 2 \pi \eta _{A}\nu _{z}Mv^{2}, \label{TEMPHEATING}
\end{equation}
where the geometrical overlap factor $\eta_A = A_b/S = 2 d /\pi R_r$ is the ratio between the
cross-sectional area of the laser beam, $A_b$, and that of the thermal cloud, $S=\pi R_r^2$.

\begin{figure}[tbp]
\begin{center}
\epsfxsize=100mm \centerline{\epsfbox{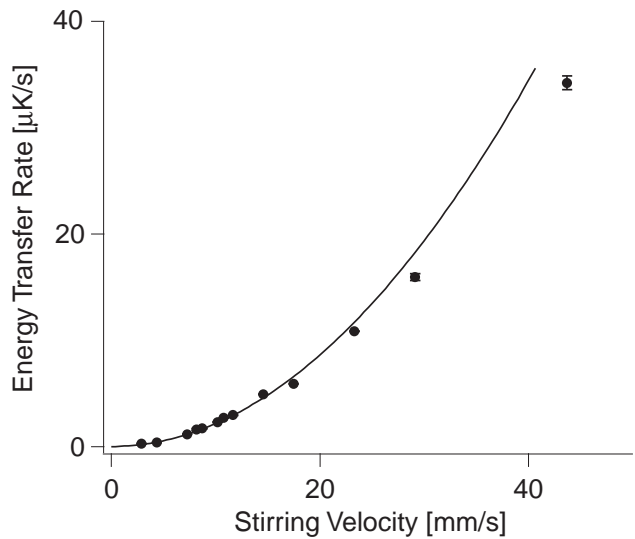}}
\vspace{0.3cm}
\end{center}
\caption{Heating of the normal gas.  The energy transfer rate per
particle versus velocity of the moving laser beam is compared to
the result of the kinetic model described in the text (solid
line).} \label{thermal_versus_velocity}
\end{figure}

Eqn. \ref{TEMPHEATING} has a simple interpretation: each atom
moves back and forth in the axial direction at a rate $\nu_z$,
and within each trapping period a fraction $\eta_A$ of them hit
the stirring beam acquiring an energy $\propto Mv^{2}$.  Fig.
\ref{thermal_versus_velocity} shows the energy transfer rate to a
gas of atoms slightly above the transition temperature, obtained
using the calorimetric technique described earlier.  We varied
the velocity by keeping the scan amplitude fixed to approximately
the diameter of the thermal cloud, and varying the scan
frequency.  The temperature was obtained from a fit to the wings
of the cloud \cite{kett99var}.  To compare with the model
calculations Eqn. \ref{TEMPHEATING} was evaluated for the average
of initial and final temperatures (the temperature increased by
about a factor of 2 during the stirring) using the experimental
parameters $d = 10 \mu$m, laser power 18 mW, $\nu_z = 20$ Hz. The
laser power was chosen to give a barrier height $U \approx 6
k_{B}T$, where $T$ is the gas temperature.  The data clearly show
the parabolic dependence on the laser beam velocity up to the
thermal velocity $V = \sqrt{2k_B T/M} = 40$ mm/s, as predicted
from Eqn. \ref{TEMPHEATING}, thus verifying the basic tenets of
the kinetic model of heating.

\begin{figure}[tbp]
\begin{center}
\epsfxsize=85mm \centerline{\epsfbox{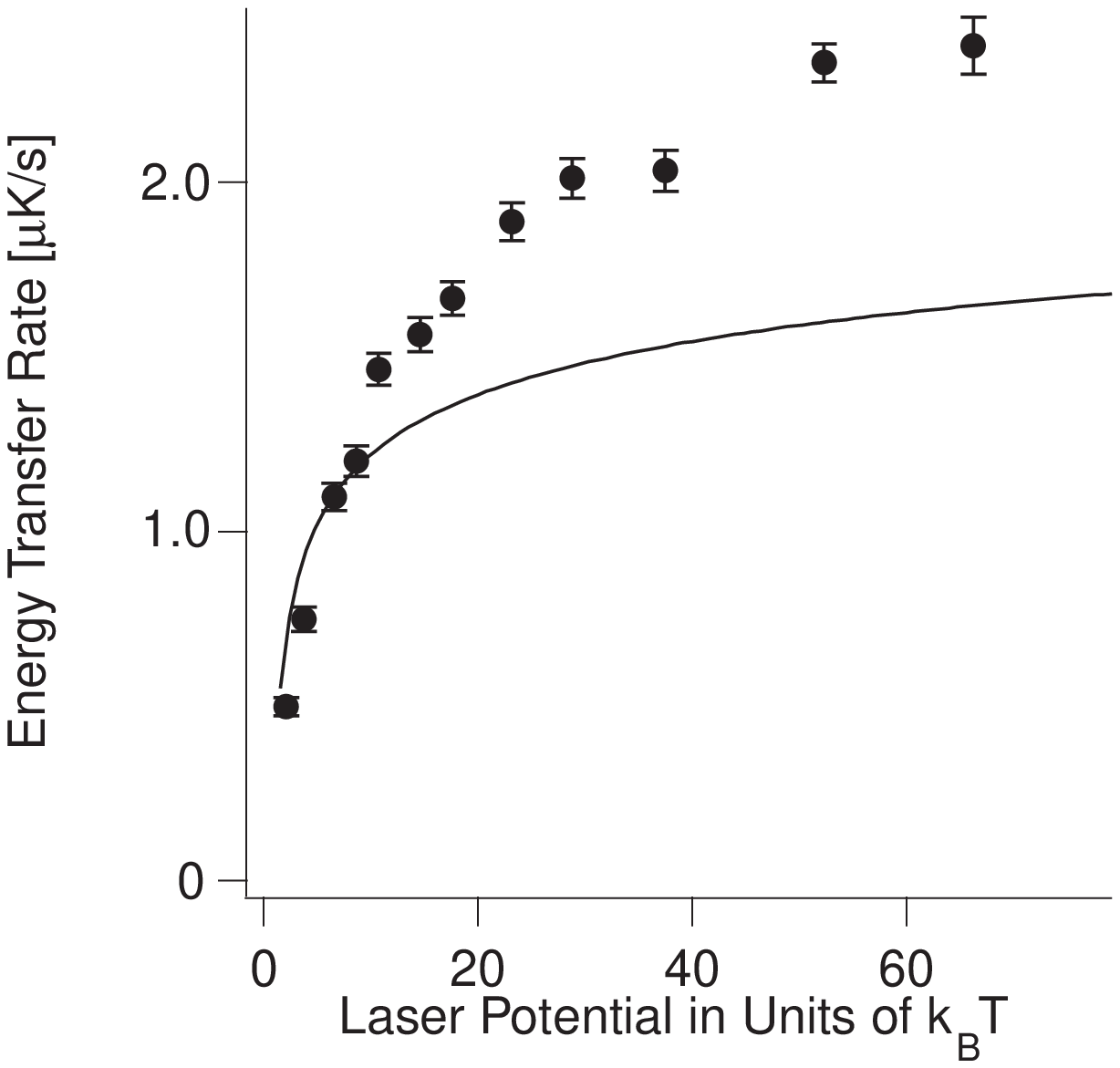}} \vspace{0.3cm}
\end{center}
\caption{Dependence of the heating on the power of the stirring
laser beam.  Shown is the heating of a thermal cloud close to the
BEC transition temperature versus the laser beam potential $U_0$
in units of the temperature $k_B T$. Data were taken by varying
the laser power at a fixed stirring velocity of 7mm/s. The solid
line is a fit to the data at low laser power $U\leq 10 k_B T$.}
\label{thermal_versus_power}
\end{figure}

The observed absolute heating rate also agreed well with the prediction given by Eqn.
\ref{TEMPHEATING} to within 20-30\%. The major uncertainty came from the laser beam size, which may
have had a 20-30\% variability from run to run.  In Fig. \ref{thermal_versus_power} we show the
heating rate for a cloud near $0.78 \mu$K, plotted versus the ratio of barrier height $U_0$ to
temperature $k_B \bar{T}$, where $\bar{T}$ is the average of initial and final temperatures.  In
this regime, the heating rate increased in proportion to the effective laser beam size, which
scales as $\sqrt{\ln{(U_{0}/k_B T)}}$. The solid line is a fit to this functional form. At higher
laser power the data deviate from this fit probably due to additional heating from the
non-Gaussian wings of the laser beam.

\begin{figure}[tbp]
\begin{center}
\epsfxsize=85mm \centerline{\epsfbox{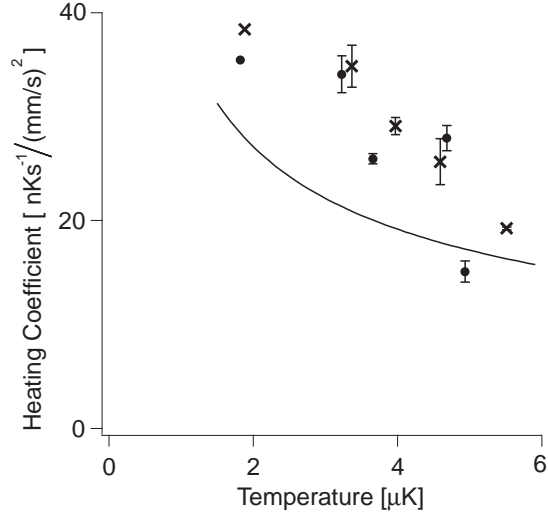}}
\vspace{0.3cm}
\end{center}
\caption{Temperature dependence of the heating.  Plotted is the heating coefficient $\kappa$
described in the text, for two different velocities, 9.7 mm/s ($\times$) and 19.4 mm/s
($\bullet$), versus the average temperature of the gas. The solid line is the value predicted by
Eqn. \ref{TEMPHEATING}. The decrease in heating efficiency versus temperature indicates decreasing
overlap between the stirrer and the gas.}\label{thermal_versus_temperature}
\end{figure}

In the above model, the temperature of the gas only appears through the factor $\eta_{A}$.
Therefore, at higher temperatures, where the cloud is more spatially extended, one expects the
heating rate to decrease as the size $R_{r}$ increases, or ${\left .\frac{dE}{dt}\right|}_N =
\kappa v^{2}$, where $\kappa \propto d/\sqrt{T}$.  We keep the ratio of barrier height to
temperature, and therefore the effective laser beam size $d$, approximately constant.  Fig.
\ref{thermal_versus_temperature} shows the coefficient of heating $\kappa$ measured for two
different stirring velocities $v$ below the thermal velocity, along with the model prediction. The
data show decreasing efficiency of heating at high temperatures, which we attribute to the decrease
of geometrical overlap between the cloud and the laser beam.

\begin{figure}[tbp]
\begin{center}
\epsfxsize=100mm \centerline{\epsfbox{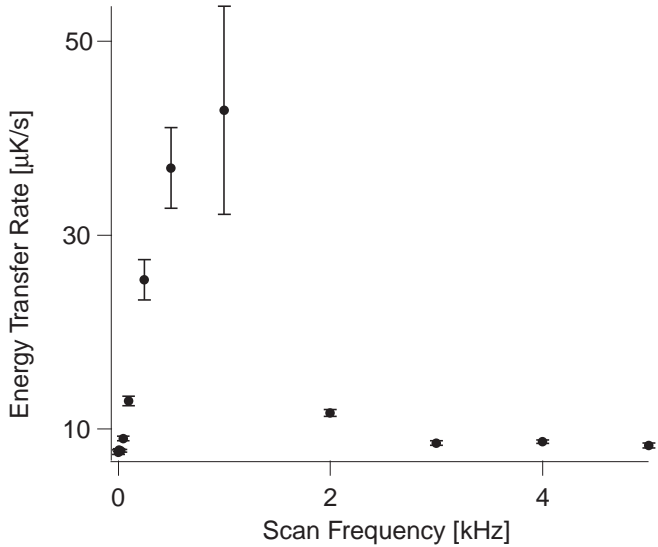}}
\vspace{0.3cm}
\end{center}
\caption{Heating of a thermal cloud close to the BEC transition temperature
versus stirring frequency.}
\label{thermal_versus_frequency}
\end{figure}

One may ask if the heating measured depends strictly on velocity,
or if there is an additional dependence on the scan frequency.
For low scan frequencies $f <20 \nu_z$, we observed no difference
in the heating for different frequencies at the same laser beam
velocity, taking care to avoid the trap resonance at $f = \nu_z$.
The data presented thus far depended only on the laser beam
velocity. However, at very high scan frequencies $f \geq 50
\nu_{z}$, the back-and-forth motion of the optical potential
becomes time-averaged and appears as a quasi-static perturbation
to the atoms. Therefore the heating rate diminished, as seen in
Fig. \ref{thermal_versus_frequency}, where the heating rate was
measured at a fixed scan amplitude and variable frequency.

\section{DISCUSSION}

Since our first report on critical velocities in a condensate,
several theoretical papers have added further insight
\cite{jack00,nore00,cres00,fedi00vcrit,stie00,wini00drag}. The
observed critical velocity $v_c \simeq 0.1 c_s$ is lower than the
predictions for homogeneous \cite{fris92,huep97} and
inhomogeneous \cite{wini99} 2D systems. This discrepancy is most
likely due to the actual 3D geometry, where the laser beam
pierced lower density regions of the condensate.  It was noted in
Ref. \cite{fedi00vcrit} that the critical velocity for phonon
excitation is lowered by the inhomogeneous density distribution.
Vortex stretching and half-ring vortices can lower the 3D
critical velocity below the 2D value \cite{nore00}. Jackson {\em
et al.} \cite{jack00,wini00drag} performed 3D simulations
obtaining a critical velocity as low as $0.13 c_s$, quite close to
our results.

The relevant critical velocity  is most likely related to vortex
{\it nucleation}
\cite{fris92,jack00,nore00,stie00,wini00drag,huep97,wini99,cara99},
which is usually smaller than the Landau and Feynman critical
velocities \cite{wilk87} at which phonons
\cite{land41,fedi00vcrit} or vortices \cite{feyn55,cres00} become
energetically favorable. In the trapped quantum gases, due to the
isolation from the environment there is generally no initial
vorticity present, and vortex nucleation, rather than the
energetics of vortex formation, plays the dominant role in
determining the critical velocity.  This is different from the
case of liquid $^4$He, where there is usually vorticity present
at the surface. As predicted by the Feynman criterion, observed
critical velocities depend strongly on the size of the flow
channel \cite{wilk87} which reflects the dependence of the vortex
energy on the finite size geometry.

There are differences between a constantly moving object and our scanning laser beam.  The rapid
turnaround of the laser beam occurring twice per cycle can emit phonons causing additional heating
\cite{jack00,wini00drag}.  This heating should scale with $v^2 f$, where $v$ is the velocity of
the scan and $f$ the scan frequency.  Our new high sensitivity measurement technique could be used
to explore this heating mechanism.

We may compare the superfluid heating with that of the normal
gas. Clearly, different mechanisms have been outlined for the
thermal cloud and for the condensate.  For a thermal gas, the
heating rate is

\begin{equation}
{\left. \frac{dE}{dt} \right|}_T = N_T \kappa_T v^2
\end{equation}
while the condensate heating rate above the critical velocity is
\begin{equation}
{\left. \frac{dE}{dt}\right |}_C = N_C\kappa_C v(v-v_c) \approx N_C \kappa_{C} v^{2}
\end{equation}
where the approximation holds for $v \gg v_c$.  In general we may
write the heating rate as $\Delta \cdot \Gamma$, where $\Delta$
is the energy per collision with the laser beam, and $\Gamma
\simeq n A V$ is the rate of such collisions.  For a thermal gas,
$V$ is the thermal velocity $V_T = \sqrt{2k_B T/M}$.  We rewrite
this as $V=\omega _z R_z$ and use this relation to obtain the
velocity $V$ for the condensate as $V_C= \sqrt{2}c_s$, where
$R_z$ is the Thomas-Fermi radius.  This could also be derived
from the pressure $n M c_s^2/2$ exerted by the condensate on the
stirrer \cite{dalf99rmp}. Therefore, in comparing heating in a
thermal cloud and a condensate we should assume that the
condensate atoms strike the stirrer with a velocity $V_C \propto
c_s$. The ratio $r$ of the energy transferred per collision with
the laser beam for normal and condensed gases is then
\begin{equation}
r = \frac{\Delta_T}{\Delta_C} = \frac{N_T \kappa_T}{N_C \kappa_C} \cdot \frac{n_C A_C V_C}{n_T A_T
V_T} \approx  \frac{\kappa_T R_T}{\kappa_C R_C}
\end{equation}
where $R_T$ and $R_C$ are the transverse sizes of the normal and
condensed components, respectively.  We measured this ratio as $r
\simeq 2$.  Thus outside the superfluid regime, the difference in
intrinsic heating for normal and superfluid components is seen to
be quite small. Each collision with the stirrer transfers an
energy of about $Mv^2$ to the cloud, both in the thermal cloud
and the condensate. Thus the observed difference between the
heating coefficients $\kappa_T$ and $\kappa_C$ is mainly
geometrical due to the different cloud sizes.

In conclusion, we have studied dissipation induced by a macroscopic stirrer in a Bose gas below
and above the BEC transition temperature.  We present high sensitivity calorimetric measurements
showing that the critical velocity is lower than suggested by earlier data\cite{rama99}.
Measurements on dilute thermal clouds support a kinetic model of heating. This provides a first
approximation to dissipation below the BEC transition temperature where one may separately
consider the contributions from condensed and non-condensed phases.

\section*{ACKNOWLEDGMENTS}

We thank A. P. Chikkatur and A. G\"{o}rlitz for experimental
assistance, and S. Rica for useful discussions. This research is
supported by NSF, ONR, ARO, NASA, and the David and Lucile Packard
Foundation

\bigskip


\end{document}